\documentclass{appolb}
\usepackage{graphicx}
\usepackage{multicol}
\usepackage{amsmath}
\usepackage{hyperref}

\begin{document}

\newcommand{\AGeVc}{$A$~GeV/$c$}
\newcommand{\AGeV}{$A$~GeV}
\newcommand{\A}{$A$}
\newcommand{\SHINE}{NA61/SHINE}

\title{New measurement of pion directed flow relative to the spectator plane by the NA49 experiment at CERN  %
\thanks{XIII Workshop on Particle Correlations and Femtoscopy (WPCF 2018)}%
}

\author{E. Kashirin$^{\mathrm{a}}$, O. Golosov$^{\mathrm{a}}$, V. Klochkov $^{\mathrm{b,c}}$, I. Selyuzhenkov $^{\mathrm{a, b}}$\\
for the NA49 collaboration
\address{$^{\mathrm{a}}$ National Research Nuclear University (Moscow Engineering Physics Institute) Moscow, Russia}
\address{$^{\mathrm{b}}$ GSI Helmholtzzentrum f\"ur Schwerionenforschung, Darmstadt, Germany}
\address{$^{\mathrm{c}}$ Goethe-University Frankfurt, Frankfurt, Germany}
}
\maketitle
\begin{abstract}
We report a new measurement of negatively charged pion directed flow $v_1$ relative to the spectator plane for Pb+Pb collisions at the beam energy 40\AGeV~recorded by the NA49 experiment at CERN. $v_1$ is reported as a function of rapidity and transverse momentum in different classes of collision centrality.
The projectile spectator plane is estimated using transverse segmentation of the NA49 forward hadron calorimeter.
The new results extend the NA49 data for $v_1$, which was previously measured only relative to the participant plane, and complement recent preliminary data by the \SHINE~ collaboration and published results from the STAR at RHIC beam energy scan program.
\end{abstract}

\section{The NA49 experiment and data sample}

The NA49 experiment at CERN SPS is a predecessor of the currently operating fixed target \SHINE~experiment which has recently extended its program for Pb-ion beam momentum scan at 13\A, 30\A~and 150\AGeVc.
NA49 have collected data for Pb+Pb collisions at beam energies of 20\A, 30\A, 40\A, 80\A { and 158\AGeV}~\cite{Alt:2003ab}.
The NA49 and \SHINE~data complements the measurement of flow harmonics available from Beam Energy Scan (BES) program of STAR at  RHIC~\cite{Adamczyk:2014ipa} and extends the measurement towards forward rapidity up to the region where projectile spectators appear. 
The new results are also relevant for studies in the few GeV collision energy range, in particular for the preparation of the Compressed Baryonic Matter (CBM) heavy-ion experiment at the future FAIR facility in Darmstadt.

A sample of Pb+Pb collisions at 40\AGeV~recorded in 2000 with minimal bias and central trigger was used in present analysis.
About 340k (440k) minimal bias (central) events with fitted vertex position close to the target were selected.

\begin{figure}[htb]
\centerline{%
\includegraphics[width=0.99\textwidth]{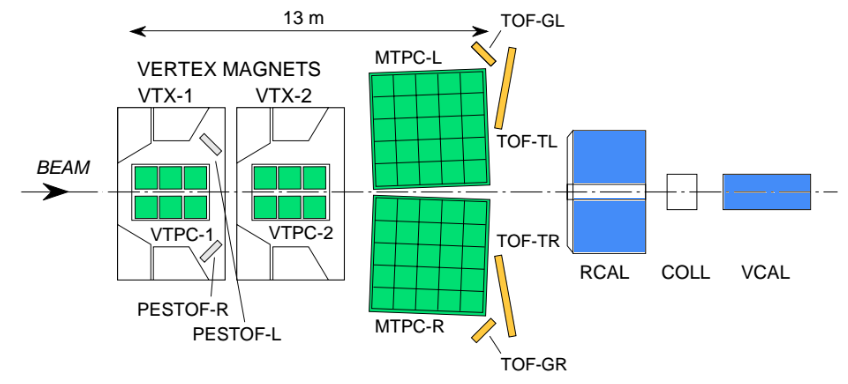}}
\caption{Schematic layout of the NA49 experiment. The four large volume time-projection chambers (TPC) are shown in green, while the Ring (RCAL) and Veto (VCAL) calorimeters are marked in blue.}
\label{Fig:NA49layout}
\end{figure}

The layout of the NA49 experiment is shown in Fig.~\ref{Fig:NA49layout}.
The NA49 experiment~\cite{AFANASIEV1999210} consists of four large volume time-projection chambers (TPC). Two vertex TPCs (VTPC-1 and VTPC-2) are positioned inside the magnet and used for momentum reconstruction. Two main TPCs (MTPC-L and MTPC-R) are used for particle identification via measurement of the specific energy loss ($dE/dx$).

\begin{figure}[htb]
\noindent
\includegraphics[width=0.49\textwidth]{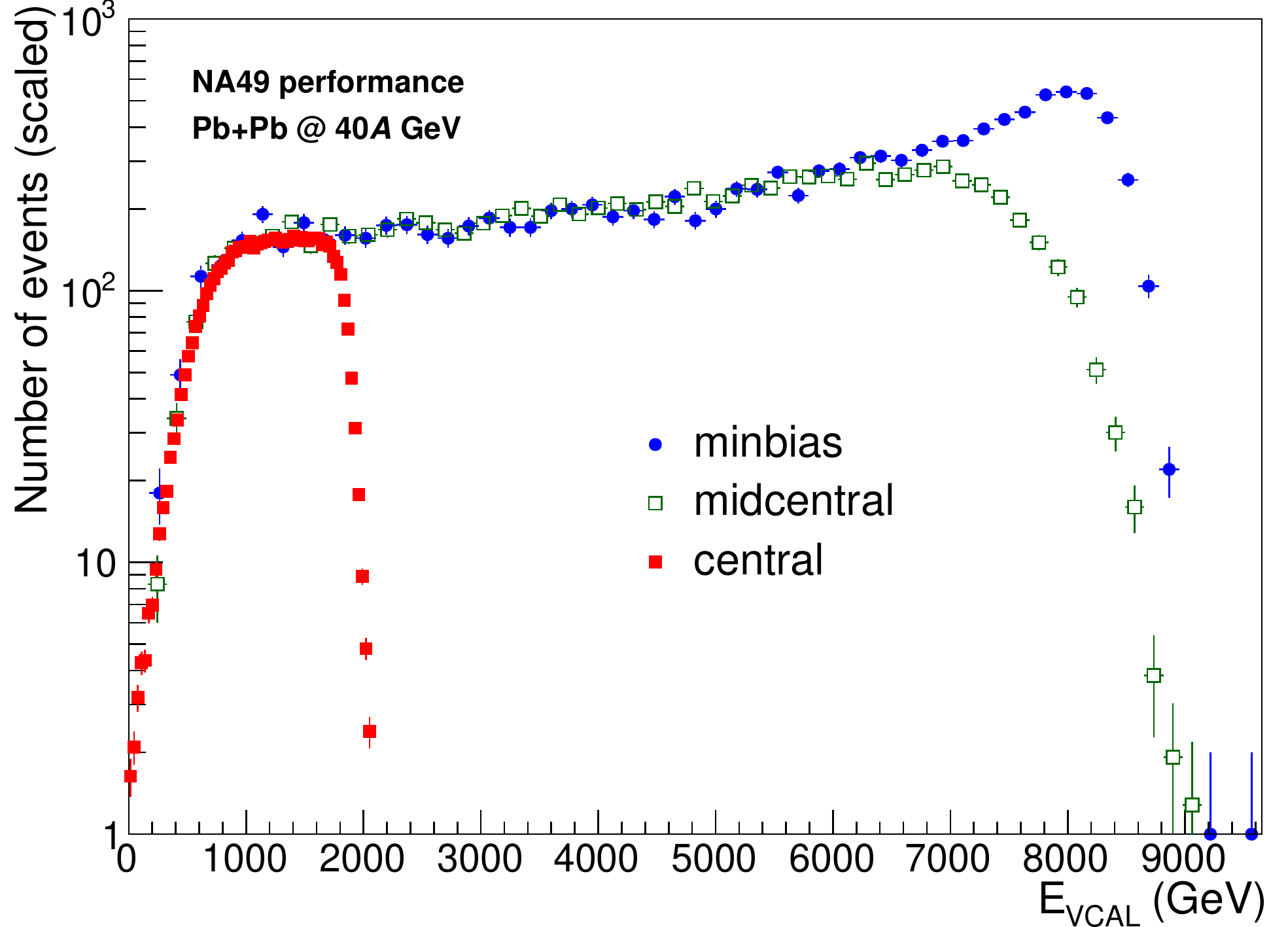}
\includegraphics[width=0.49\textwidth]{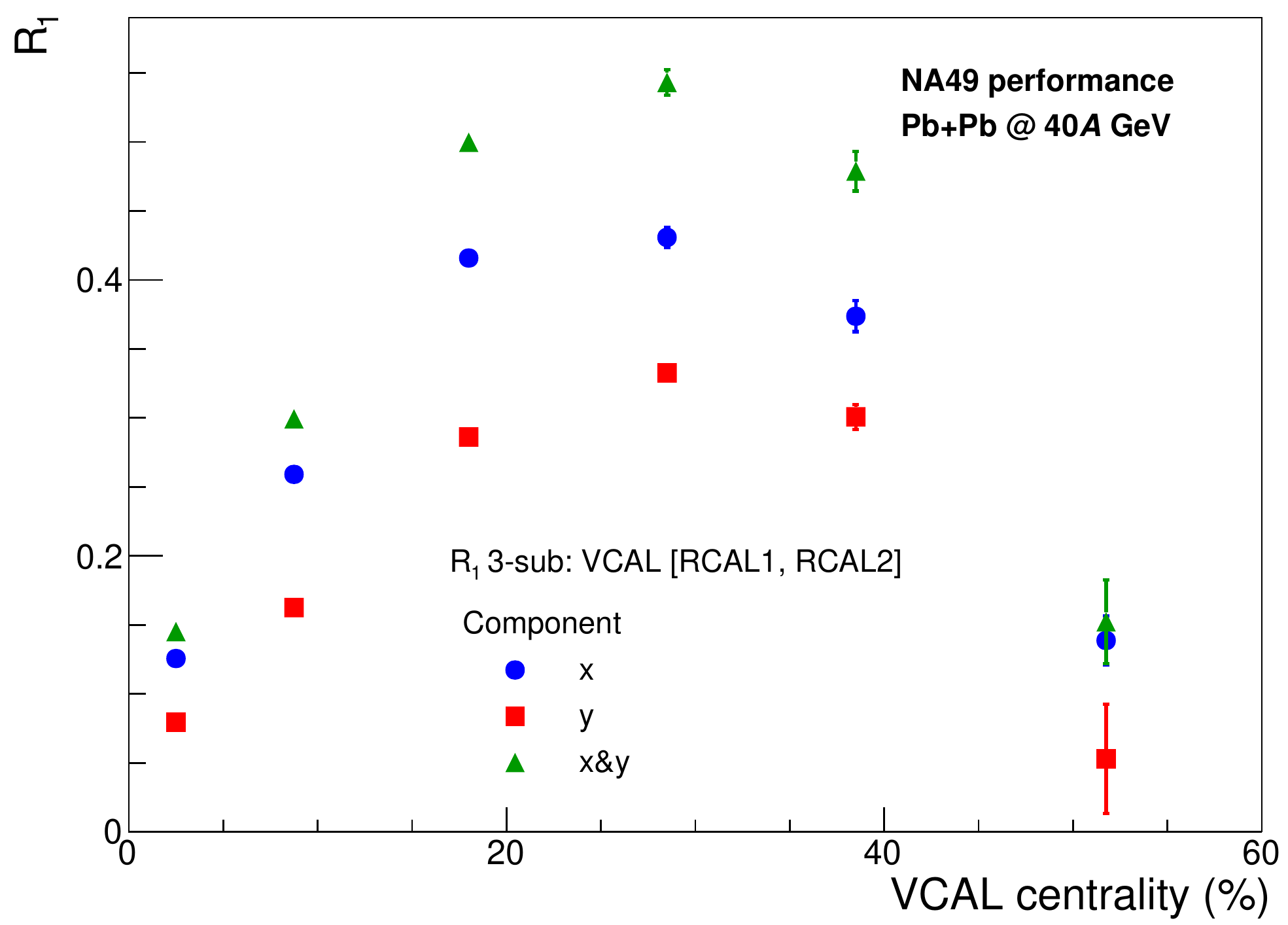}
\caption{
(left) VCAL energy ($E_{\rm VCAL}$) distribution for different triggers.
(right) Spectator plane resolution correction factor ($R_1$) for VCAL vs. collision centrality. Calculations for $x$ and $y$ $Q$-vector components separately and combined are shown.}
\label{Fig:v1pt}
\end{figure}

Veto (VCAL) calorimeter is installed 20 meters downstream the target behind the collimator and has a 2x2 transverse module granularity. The opening of the collimator is adjusted such that beam particles, projectile fragments and spectator neutrons and protons can reach the calorimeter.
Deposited energy in VCAL modules is sensitive mostly to projectile spectators and is used for centrality and spectator plane estimation.
Figure \ref{Fig:v1pt}(left) shows the distribution of energy in VCAL for different centrality triggers.
Ring (RCAL) calorimeter is positioned at 18 meters from the target and its transverse granularity of 10 rings with 24 modules each is used to estimate the spectator plane resolution of VCAL. Figure \ref{Fig:v1pt}(right) shows the extracted spectator plane resolution correction factor of VCAL from the correlation with RCAL signals (more details in Sec.~2 below).

Only tracks with at least 20 points in VTPCs and and at least 30 points in all TPCs were considered.
To avoid track splitting the number of hits associated with a track was required to be more than 55\% of total hits possible.
Primary tracks were selected using distance of closest approach (DCA) to the fitted primary vertex.
Only tracks with DCA less than $2 ~cm$ in magnetic field bending ($x$) direction and less than $1 ~cm$ in a perpendicular ($y$) direction were considered.

\section{Flow analysis details}

Directed flow $v_1$ is measured from correlation of two-dimensional flow vectors $\mathbf{q_1}$ and $\mathbf{Q_1}$. The $\mathbf{q_1}$ is calculated event-by-event from azimuthal angles $\phi_i$ of the i-th particle's momentum:
\begin{equation}
    	\label{Eq:q1_TPC}
		\mathbf{q_1} = \frac{1}{M_{\pi^{{}-{}}}} \sum_{i=1}^{M_{\pi^{{}-{}}}} \mathbf{u_{1,i}},
\end{equation}
where $\mathbf{u_{1,i}} = ( \cos{\phi_i}, \sin{\phi_i})$ and $M_{\pi^{{}-{}}}$ is negatively charged pion multiplicity in a given $p_T$ and rapidity ($y$) range.
Spectators' symmetry plane is estimated with the $\mathbf{Q_1}$ direction determined from azimuthal asymmetry of energy deposition in VCAL:
\begin{equation}
	\mathbf{Q_1^{\rm VCAL}} = \frac{1}{E_{\rm VCAL}} \sum_{i=1}^{4} E_i \mathbf{n_i}~,
\label{Eq:Q_VCAL}
\end{equation}
where $E_{\rm VCAL} = \sum_{i=1}^{4} E_i$ is the total energy deposited in VCAL. Unit vector $\mathbf{n_i}$ points in the direction of the center of i-th VCAL module and $E_i$ is the energy deposited in it.
Similar equation to Eq.~(\ref{Eq:Q_VCAL}) is used to define the $Q_1^{\rm RCAL}$ from RCAL modules. For this the RCAL module rings are divided into two subgroups: 5 inner (RCAL1) and 5 outer (RCAL2) rings.

Independent estimates of directed flow $v_1$ are obtained using the scalar product method:
\begin{equation}
	v_1^\alpha \lbrace A|B,C \rbrace = \frac{2 \langle q_{1, \alpha} Q^A_{1,\alpha} \rangle}{R^A_{1,\alpha} \lbrace B, C \rbrace},
\end{equation}
where $\alpha = x, y$ are $q_1$ and $Q_1$ components and $A$=VCAL, $B$=RCAL1, and $C$=RCAL2. Correction factors $R^A_{1,\alpha} \lbrace B, C \rbrace$ were calculated with three-subevent method:
\begin{equation}
R^A_{1,\alpha} \lbrace B, C \rbrace = \sqrt{2 \frac{
	\langle Q^A_{1, \alpha} Q^B_{1, \alpha} \rangle \langle Q^A_{1, \alpha} Q^C_{1, \alpha} \rangle 
    }{
    \langle Q^B_{1, \alpha} Q^C_{1, \alpha} \rangle
    }}.
\end{equation}
Effects of transverse momentum and rapidity dependent tracking efficiency in $v_1$ measurement were corrected using Monte-Carlo simulation of the NA49 detector response.
For this a sample of heavy-ion collisions generated using VENUS event generator~\cite{Werner:1993uh} was used. Non-uniform azimuthal acceptance of the NA49 experiment was corrected for using the procedure described in~\cite{Selyuzhenkov:2007zi} and implemented in the QnCorrections framework~\cite{QnGithub:2015,Gonzalez:2016GSI}. Recentering, twist and rescale corrections were applied as a function of time and centrality.

\section{Results}

Results are reported for negatively charged pions ($\pi^{-}$) produced by strong interaction process and weak decays (within TPC acceptance).
\begin{figure}[htb]
\noindent
\includegraphics[width=0.49\textwidth]{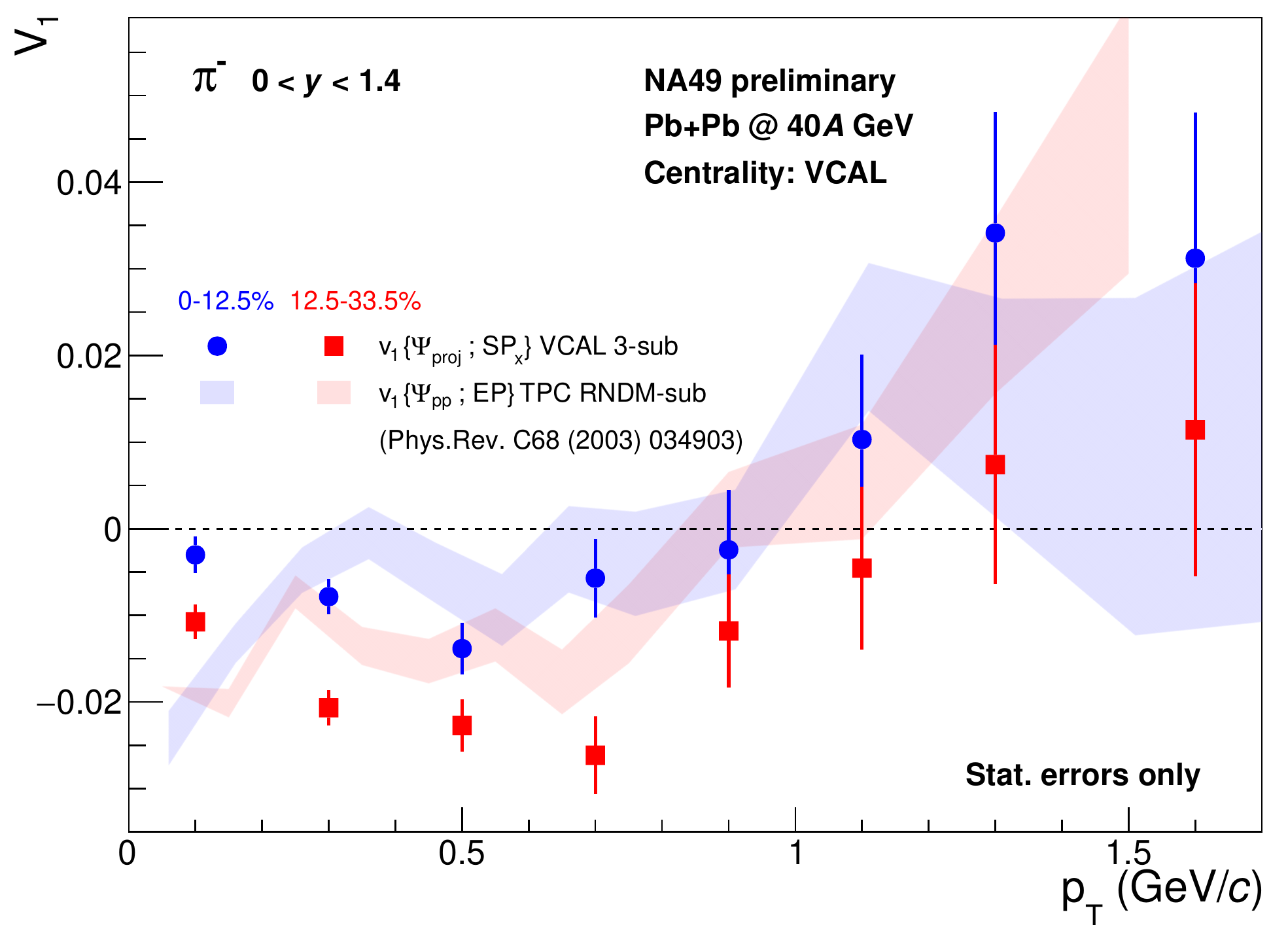}
\includegraphics[width=0.49\textwidth]{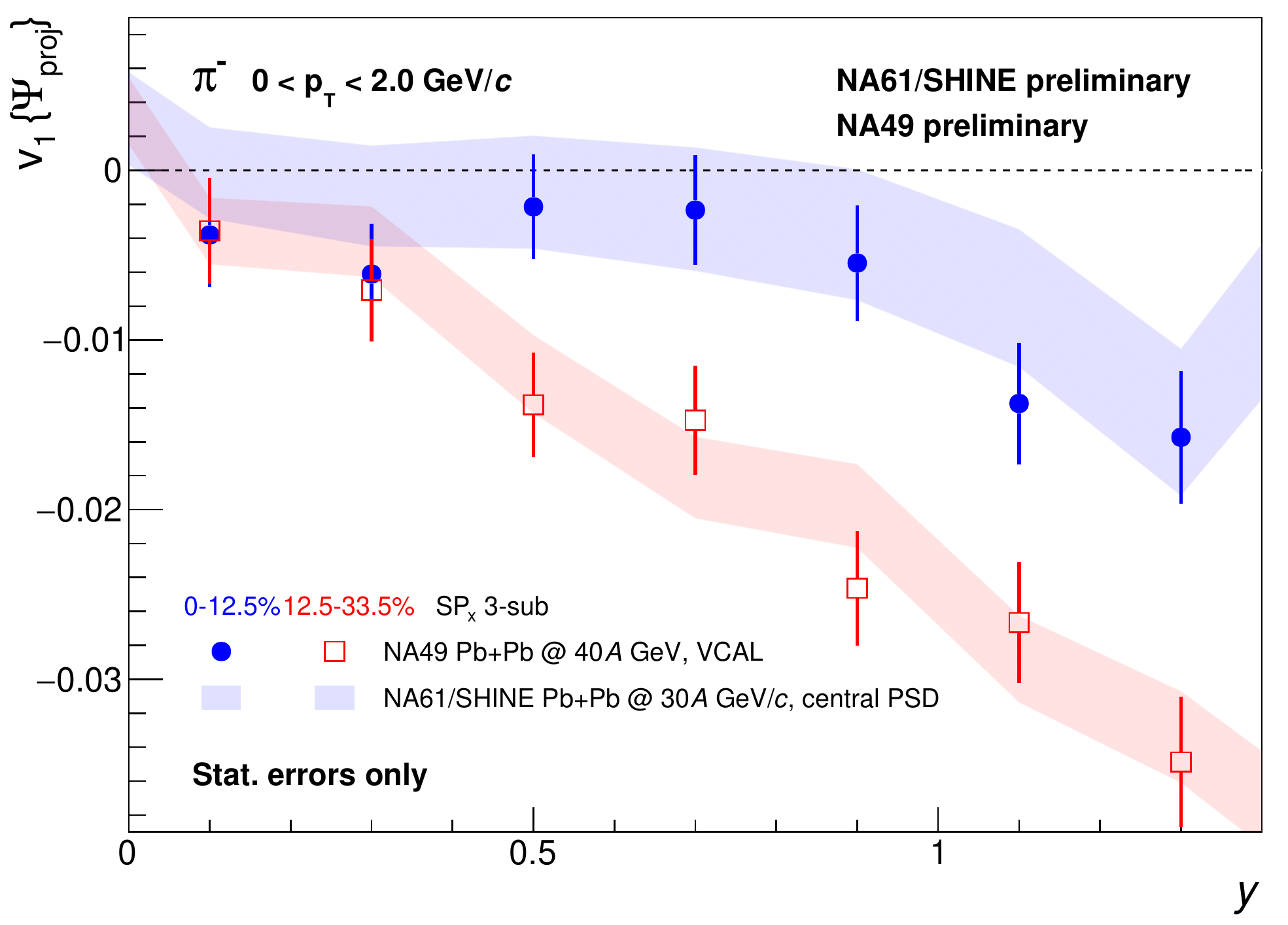}
\caption{Directed flow ($v_1$) of negatively charged pions in different centrality classes.
(left) $v_1$ vs. transverse momentum ($p_T$). (right) $v_1$ vs. rapidity ($y$).
Only statistical errors are shown.}
\label{Fig:v1pt_comp}
\end{figure}
Corrections for detector acceptance non-uniformity is applied according to description in Sec.~2 above.
Independent $v_1$ estimates with $x$ and $y$ Q-vector components are found to be consistent within statistical uncertainties.
Due to the azimuthal TPC acceptance, the correlation for $x$ components are better defined statistically and thus are used for presenting new NA49 preliminary results.

Figure~\ref{Fig:v1pt_comp}(left) shows NA49 results for directed flow $v_1$ vs. transverse momentum ($p_T$) measured relative to the projectile spectator plane in Pb+Pb collisions at 40\AGeV. 
Results are compared to previously published NA49 results for the directed flow relative to the participants plane. Different sensitivity of the directed flow wrt. the participant and spectator planes are observed.  

Figure~\ref{Fig:v1pt_comp}(right) shows results for $v_1$ vs. rapidity in Pb+Pb collisions at 40\AGeV. 
Results are compared to the new preliminary data\cite{qm_proceedings} for directed flow relative to spectator plane in Pb+Pb collisions at 30\AGeV/$c$ reported recently by the \SHINE~Collaboration. Despite small difference in the collision energy, a good agreement between $v_1$ from NA49 and \SHINE~Collaboration are observed.

\section{Summary and Outlook}

Preliminary results for directed flow of negatively charged pions relative to the projectile spectator plane in Pb+Pb collisions at 40\AGeV~are presented for different centrality classes as a function of transverse momentum and rapidity. Good agreement of directed flow estimated with new technique is shown for the energies 30\AGeV~(\SHINE) and 40\AGeV~(NA49).

For Pb+Pb at 40\AGeV~it is planned to study elliptic flow and make systematics studies. The new developed technique of azimuthal flow measurement relative to spectator plane will also be applied to other energies available at NA49.

\section{Acknowledgements}

This work was partially supported by the Ministry of Science and Education of the Russian Federation, grant N 3.3380.2017/4.6, and by the National Research Nuclear University MEPhI in the framework of the Russian Academic Excellence Project (contract No. 02.a03.21.0005, 27.08.2013).

\bibliography{references}
\bibliographystyle{ieeetr}

\end{document}